\documentclass[referee]{aastex}

\slugcomment{to be published in The Astrophysical Journal}

\begin{document}

\title{Membership and Multiplicity among Very Low-Mass Stars and Brown 
Dwarfs in the Pleiades Cluster\altaffiltext{1}{Based 
in part on observations made with the NASA/ESA Hubble Space
Telescope, obtained at the Space Telescope Science Institute, which is operated
by the Association of Universities for Research in Astronomy, Inc., under NASA
contract NAS 5-26555. These observations are associated with proposal ID 7952.}}

\author{E. L. Mart\'\i n}
\affil{Division of Geological and Planetary Sciences, California Institute of Technology 150-21, Pasadena, CA 91125, USA, ege@gps.caltech.edu}
\affil{Astronomy Department, University of California at Berkeley, 601 Campbell Hall, Berkeley, CA 94720, USA}

\author{W. Brandner}
\affil{Institute for Astronomy, University of Hawaii, 
2680 Woodlawn Drive, Honolulu, HI 96822, USA, brandner@nic.ifa.hawaii.edu}

\author{J. Bouvier}
\affil{Observatoire de Grenoble, B.P.53, F-38041 Grenoble Cedex 9, France, 
Jerome.Bouvier@obs.ujf-grenoble.fr}

\author{K. L. Luhman  \& J. Stauffer}
\affil{Smithsonian Astrophysical Observatory, 60 Garden St., Cambridge, 
MA 02138, USA, kluhman@jayhawk.harvard.edu, stauffer@amber.harvard.edu}

\author{G. Basri}
\affil{Astronomy Department, University of California at Berkeley, 601 Campbell Hall, Berkeley, CA 94720, USA, basri@soleil.berkeley.edu}

\and 

\author{M. R. Zapatero Osorio}
\affil{Instituto de Astrof\'\i sica de Canarias, 38200 La Laguna, 
Tenerife, Spain, mosorio@ll.iac.es}

\newpage

\begin{abstract}

We present near-infrared photometry and optical spectroscopy 
of very low-mass stars and brown dwarf candidates in the Pleiades open cluster.
The membership status of these objects is assessed using 
color-magnitude diagrams, lithium and spectral types. Eight objects out of 45 
appear to be non-members. 
A search for companions among 34 very low-mass Pleiades members 
(M$\le$0.09~M$_\odot$) in high-spatial resolution images obtained with the  
Hubble Space Telescope and the adaptive optics system of the Canada-France-Hawaii 
telescope produced no resolved binaries with separations larger than 0.2 arcsec 
(a$\sim$27~AU; P$\sim$444 years). Nevertheless, we find evidence for a   
binary sequence in the color-magnitude diagrams, in agreement with the 
results of Steele \& Jameson (1995) for higher mass stars. 
We apply the lithium test to two objects: CFHT-Pl-16, which lies in the 
cluster binary sequence but is unresolved in images obtained with the 
Hubble Space Telescope;  
and CFHT-Pl-18, which is binary with 0".33 separation (Mart\'\i n et al. 
1998). The first object passes the test, but the second object does not. 
We conclude that CFHT-Pl-16 is an Pleiades brown dwarf binary with separation $<$11~AU, 
and that CFHT-Pl-18 is a foreground system. 
We compare the multiplicity statistics of the Pleiades very low-mass 
stars and brown dwarfs with that of G and K-type main sequence stars in the solar neighborhood  
(Duquennoy \& Mayor 1991). We find that there is some evidence for a  
deficiency of wide binary systems (separation $>$27 AU) among the Pleiades very low-mass 
members. 
We briefly discuss how this result can fit with current scenarios of brown dwarf formation.  
We correct the Pleiades substellar mass function for the 
contamination of cluster non-members found in this work. 
We find a contamination level of 33\% among the brown dwarf candidates 
identified by Bouvier et al. (1998). Assuming a power law IMF across the 
substellar boundary, 
we find a slope dN/dM$\sim$M$^{-0.53}$, implying that the number of 
objects per mass bin is still rising but the contribution to the total mass 
of the cluster is declining in the brown dwarf regime.

\end{abstract}


\newpage 

\section{Introduction}

Brown dwarfs (BD) provide the opportunity of extending classical studies of stellar 
properties into the hitherto unexplored realm of objects with 
masses lower than the H-burning limit (0.075~M$_\odot$; e.g. Baraffe et al. 1998). 
Stars are frequently associated 
with other stars in multiple systems. 57\% of nearby G-type dwarfs have H-burning 
companions (Abt \& Levy 1976; Duquennoy \& Mayor 1991, hereafter DM91). The multiplicity frequency 
seems to be lower among nearby M-type dwarfs (38\%, Henry \& McCarthy 1990; Fisher \& Marcy 1992). 
This difference could be due to the smaller mass range available for the stellar secondaries 
of M dwarfs. On the other hand, the distribution of binary separations is similar for 
G- and M-type stars. It has a broad maximum from 3 to 30 Astronomical Units (AU). 
There is no indication that stellar binary properties depend on primary mass.  

Stellar clusters are important for studying the multiplicity properties as a function 
of age, metallicity and mass. 
One of the best studied open clusters is the Pleiades. 
It is nearby (d=125~pc), young (120~Myr), the metallicity is close to solar, and there are 
more than 800 known members (see Hambly 1998 for a review). 
The Pleiades mass function has been studied over a broad mass range,   
from 4~M$_\odot$ to 0.04~M$_\odot$ (Hambly, Hawkins \& Jameson 1993; 
Meusinger, Schilbach \& Souchay 1996; Bouvier et al. 1998, hereafter B98). 
Cluster BDs have been confirmed via the lithium test (Basri, Marcy \& Graham 1996; 
Rebolo et al. 1996). 
Several imaging surveys have recently identified a numerous population of Pleiades 
BD candidates 
(B98; Festin 1998; Zapatero Osorio et al. 1999; Hambly et al. 1999). 
Follow-up observations of these objects is necessary for assessing their membership. 

The search for BD binaries is interesting because of the following reasons: 1) 
Any fainter and cooler secondary of a BD should have even lower mass than the primary. 
2) The binary frequency among BDs and the distribution of orbital 
periods and eccentricities is an important clue for understanding 
the formation of these objects. 3) BD binaries provide the opportunity 
to measure dynamical masses, which are necessary for calibrating 
evolutionary models. 4) The Pleiades substellar 
mass function needs to be corrected for binarity. 

The paper is organized as follows: in Section 2 we present the observations and describe 
the data analysis. In Section 3 we search for companions and we construct color magnitude 
diagrams. In Section 4 we discuss the cluster membership of our objects, we derive 
parameters for the likely members, we estimate the binary frequency, and we discuss the 
implications of our results for the substellar mass function.

\section{Observations and Data Analysis}

\subsection{Sample Selection}

Our list was selected among the following Pleiades VLM candidate members 
(M$\le$0.1~M$_\odot$):  
`Calar' and `Teide' objects from Zapatero Osorio et al. 
(1997b) and Mart\'\i n et al. (1998b); `CFHT-Pl' objects from B98; 
`HHJ' objects from  Hambly et al. 
(1993); `MHO' objects from Stauffer et al.  (1998b); `PPl' objects from 
Stauffer et al. (1989, 1994) and `Roque' objects
from Zapatero Osorio et al. (1997a, 1997c, 1999). These surveys include most of the known 
Pleiades BD candidates. 

For the HST/NICMOS program, we gave preference to objects 
with known near-infrared (NIR) magnitudes and/or spectral types, and we included 
all the Pleiades BDs with lithium detections (Rebolo et al. 1996; Mart\'\i n 
et al. 1998; Stauffer et al. 1998a), with the exception of PPl~15 (Stauffer et al. 1994; 
Basri et al. 1996) because archive HST observations were already available. 
We observed a total of 30 objects, but one of them was a repetition (PPl1=Roque15). 

For the ground-based near-IR and spectroscopic observations, we selected all the CFHT objects 
that had not been observed in previous campaigns. We obtained near-IR photometry for 22 
objects and spectroscopy for 17.

\subsection{NICMOS Observations}

The targets were centered in the field of view of the  
NICMOS camera 1 (NIC1). Exposures of 447.95~s, 383.95~s and 383.95~s
were obtained in multiple-accumulate mode 
with filters F110M, F145M, and F165M (Thompson et al. 1998), respectively. 
Each target was observed during one orbit (average visibility 52 minutes). 

We used the IRAF/DIGIPHOT package for data reduction and analysis. Magnitudes for
all the targets were computed based on the header keyword PHOTFNU and are
given in Table~1.  These values should be used with caution because the values
for PHOTFNU are valid for sources with a constant flux per unit wavelength
across the band pass, which might not be the case for our sources.  Other
limiting factors in the accuracy of the magnitudes are that we used 
model point spread functions (PSF) for
fitting the data that did not always provide a perfect match to the observed PSF, 
and uncertainties in the NICMOS darks and resulting spatial variations 
in the background. 
We used on-orbits darks (as opposed to the model darks used in the standard
NICMOS pipeline) to improve the photometric accuracy, but the correction 
was still not perfect. 

\subsection{Archive HST Data}

The following Pleiades VLM objects have been observed with WFPC2 in other HST 
programs: HHJ~3, HHJ~5, HHJ~6, HHJ~10, HHJ~11, HHJ~14, HHJ~19, 
HHJ~36 and PPl~15. 
We retrieved the data from the HST archive. 
None of them shows any companion in the F785LP filter 
up to 4 magnitudes fainter 
than the primary at separations between 0".15 and 4."0. 

\subsection{Ground-based Infrared Observations}

Near-IR broad-band data was collected in the following observing runs: 
1) On 21-23 September 1997 at the 1~m Nickel telescope 
of Lick Observatory using the LIRC II camera. The wide field of view (FOV) 
was selected (7.29 arc mins$^{2}$, 1.71 arc secs pix$^{-1}$). 
2) On 14-16 November 1997 using the
1.2m telescope and the STELIRCAM IR camera. STELIRCAM
obtains J and K band data simultaneously using two 256x256
InSb detector arrays and a dichroic filter.  The pixel
scale used for these observations was 0.3 arcseconds per pixel.
The typical total integration time per object was about 
20 minutes, and the seeing was about 1.5 arcseconds on
average. 3) On 13-15 January 1998 at the 3.6~m Canada-France-Hawaii telescope 
(CFHT) using  the
adaptive optics (AO) system PUEO (Rigaut et al. 1998) and the KIR (Doyon et al. 1998) 
IR camera with a 30
arcsec FOV. Integration times ranged from 100 to 150 seconds in JHK
broad-band filters. Images were dark and flat-field corrected and aperture
photometry performed with IRAF/APPHOT. UKIRT faint standards were observed
every night for photometric calibration. For 10 of the 12 CFHT candidates
observed during this run (see table 2), the adaptive optics system provided
an angular resolution of 0.2 to 0.3 arcsec FWHM. The 2 remaining objects,
CFHT-17 and 18, had no proper AO guiding star and have 0.4-0.5 arcsec FWHM
on the final images. 

\subsection{Spectroscopy}

We obtained spectra using the Keck~II telescope with the Low Resolution Imaging Spectrograph
(LRIS; Oke et al. 1995), the 4~m telescope at Kitt Peak National 
Observatory (KPNO) with the Cryogenic Camera (Cryocam) spectrograph, and the 4.2~m 
William Herschel Telescope (WHT) with ISIS.  
The observing log is given in Table~3. We rotated the slit in order 
to be at parallactic 
angle at the beginning of the exposure in order to minimize refraction losses.    

All the CCD frames were bias subtracted, flat fielded, sky subtracted, 
and variance extracted using routines 
within IRAF\footnote {IRAF is distributed
by National Optical Astronomy Observatory, which is operated by the Association
of Universities for Research in Astronomy, Inc., under contract with the
National Science Foundation.}. 
We calibrated in wavelength using the 
emission spectrum of HgNeAr lamps. Correction for instrumental response was made using 
spectra obtained in the same night of flux standards with fluxes 
available in the IRAF environment. 
We did not use order-blocking filters for the Keck observations 
of the Pleiades BD candidates. These objects are so red that second-order contamination is negligible. 
We observed VB10 without filter and with the OG570 order-blocking filter. We did not find 
any significant contamination due to second-order light. We calibrated the LRIS spectra with 
the flux standard star BD+262606, which was observed the same night as the Pleiades BD candidates 
and with same instrumental configuration.  
The spectrum of the standard star obtained without filter was 
used blueward of 680~nm, and that obtained with the OG570 filter was 
used redward of 680~nm. At KPNO we used the OG550 filter to block second-order light. 
The Cryocam and WHT spectra were flux calibrated with standard stars that have 
data available in the IRAF database.

\section{Results}

\subsection{NICMOS Color-Magnitude Diagram}

Fig.~1 shows a color-magnitude diagram (CMD hereafter)                      
in the NICMOS F110M and F165M filters. 
The F145M filter is strongly affected by water vapour in the spectrum of 
very cool objects. The comparison with 
the theoretical models is less reliable because of uncertainties in the 
steam line lists and opacities. 
Our objects are plotted in the CMD together with the theoretical isochrones (age 120 Myr) 
of Chabrier et al. (2000). 
The solid line denotes the dust-free models (Nextgen; Allard et al. 1997; 
Hauschildt et al. 1999). 
The dashed line represents the new models that include dust effects (Dusty; 
Allard et al. 1998). It is expected that for temperatures cooler 
than 2800~K dust forms in the atmosphere of ultracool dwarfs (Jomes \& Tsuji 1997; 
Marley et al. 1999). In the Nextgen models, 
a T$_{\rm eff}$ of 2800~K corresponds to a mass of 0.075~M$_\odot$ and M$_{F110M}$=10.31. 
The effect of dust formation in Fig.~1 is to make the objects bluer than the Nextgen isochrone 
because of ``backwarming'' (e.g. Leggett, Allard \& Hauschildt 1998). 
We start seeing clear evidence for a turnover 
to bluer colors in the Pleiades sequence at M$_{F110M}\sim$11.5, which corresponds to 
a T$_{\rm eff}$=2320~K. Hence, we find that for T$_{\rm eff}$ lower than about 2300~K, 
dust becomes an important opacity source in the atmospheres of Pleiades BDs. 
The Dusty isochrone gives a good fit to the 
position of our faintest object (Roque~25, SpT=L0). Intermediate models between 
Nextgen and Dusty are not available. They seem to be required to fit the location of Pleiades 
BDs between M$_{F110M}\sim$11.5 and M$_{F110M}\sim$12. 

All of our targets with previously known 
lithium detections are located close (F110M-F165M=$\pm$0.2 mag) 
to the Nextgen isochrone in Fig.~1. 
These lithium BDs are bona-fide Pleiades members and help to define the cluste sequence. 
They are: 
CFHT-Pl~15, MHO~3, PPl~1, Roque~13, Teide~1 and Teide~2. 
Calar~3 has not been plotted because the photometry 
is very uncertain due to guiding problems. 

PPl~1 was observed with NIC1 at two different 
epochs (1 Sep 1998 and 9 Sep 1998) because we had not realized that it is actually  
the same object as Roque~15 
(Zapatero Osorio et al. 1997c did not notice it neither). 
We obtained different magnitudes for the two sets of observations (Table~1). 
In the second epoch the object was brighter and bluer. We have marked the variation 
in the location of PPl~1 in the CMD of Fig.~1. 
The difference is much larger in the F145M filter (8~$\sigma$) than in the other 
two filters, suggesting variability in the steam absorption. This could be a hint 
of a ``weather'' change in the brown dwarf.  Periodic photometric variability has 
been observed in I-band CCD observations of two cluster VLM stars, one in  
 $\alpha$Per (Mart\'\i n \& Zapatero Osorio 1997) and the other   
in the Pleiades (Terndrup et al. 1999). 
The light changes in these stars are probably due to surface thermal 
inhomogeneities (dark spots) modulated by the rotation of the object(s). 
On the other hand, Bailer-Jones \& Mundt (1999) have recently failed to detect 
photometric variability in Calar~3, Roque~11 and Teide~1. Further photometric 
monitoring of PPl~1 is necessary to clarify if the observed variability in NICMOS 
filters is due to weather or magnetic spots. 

The example of PPl~1 indicates that some of the spread observed in the Pleiades sequence 
of Fig.~1 could be due to intrinsic variability of the objects. Other sources of scatter 
in the CMD are unresolved binaries and non-members of the cluster. We note two examples 
of each kind of object. CFHT~Pl~16 is brighter and redder than the rest of the sequence. 
It also stands out in other diagrams that are discussed in the next sections, 
and was noted as a possible binary by B98. We consider 
it as a likely unresolved binary. CFHT~Pl~22 is much bluer and fainter than the Pleiades 
sequence. It is likely a non-member. The membership status of our sample is discussed 
in detail in Section 4.1.

\subsection{Broad-Band Color-Magnitude Diagram}

We have combined our new broad-band near-IR photometry with data available in the 
literature (B98; Festin 1998; Zapatero Osorio et al. 1997a,b; Mart\'\i n et al. 1998c) 
to produce the I vs I-K CMD shown in Fig.~2.                                    
We used the following extinction corrections 
for all the objects: A$_I$=0.06, A$_K$=0.01.  
We have compared the data with theoretical isochrones (ages=100 and 120~Myr).  
Dusty and Nextgen models are represented with dashed and solid lines, respectively. 
There is a well defined observational sequence, which is conveniently fitted by the 
the Nextgen isochrones down to M$_{\rm I}$=14. Our faintest object, Roque~25, 
is bluer than the dust-free isochrone. Its location in Fig.~3 is in good agreement 
with the Dusty isochrone. This is explained by the greenhouse effect of dust grains 
(Chabrier et al. 2000). 

The objects located well above the Nextgen isochrone could be binaries 
with nearly identical components. One of them is PPl~15, which is known 
to be a short-period nearly equal-mass binary (Basri \& Mart\'\i n 1999). 
The other objects are: NOT~1, HHJ~6, CFHT-Pl~6, CFHT-Pl~12 and 
CFHT-Pl~16. Only the later one was included in our HST/NICMOS program. HHJ~6 was observed 
with HST/WFPC2, and  CFHT-Pl~12 was observed with the CFHT AO system. If they are binaries, 
they must have angular separations smaller than 0".2 (25~AU). 

The objects located well below the Nextgen isochrone for M$_{\rm I}\ge$14 are 
likely non-members. They are identified with five-pointed star symbols in Fig.~2. 
They are CFHT-Pl 19, 20 and 22.

\subsection{Spectral Type-Magnitude Diagram}

We display the spectra of our faintest BD objects in Fig.~3. 
We derived spectral types for our targets using the calibration 
of the pseudocontinuum index 
(PC3) given by Mart\'\i n et al. (1999). 
The results are given in Table~4. 
Our values are in good agreement with previous work for the objects in common. 
Using these spectral types and those published by Steele \& Jameson (1995), 
Mart\'\i n, Rebolo \& Zapatero Osorio (1996), Cossburn et al. (1997), 
Zapatero Osorio et al. (1997c) and Festin (1998), we have made the diagram shown in Fig.~4. 
Two objects lie well outside the Pleiades sequence, one above it (CFHT-Pl~16) and one below it 
(CFHT-Pl~26). The first one is likely an unresolved binary, and the second one is probably 
a non-member. 

\subsection{The Lithium Test in CFHT-Pl~16 and CFHT-Pl~18}

CFHT-Pl~18 is an important object because it is the only binary that we have resolved 
in our HST images. It lies on the cluster sequence in the CMD diagrams discussed above.  
Mart\'\i n et al. (1998a) obtained a radial velocity consistent with cluster membership. 
In order to confirm its membership, 
we obtained additional mid-resolution spectra of CFHT-Pl~18 around the LiI  
resonance line at 670.8~nm for applying the lithium test for BDs (Magazz\`u et al. 1993). 
If CFHT-Pl~18 is indeed a member, both components should have 
preserved their initial lithium content because they are fainter than the Pleiades 
substellar boundary (Mart\'\i n et al. 1998b; Stauffer et al. 1998a). 
In Fig.~5, we show the final spectrum where we do not detect the lithium feature.  
We put an upper limit to the LiI equivalent width of 200 mA, which is a factor 
of 5 lower 
than the equivalent width (EW) measured in Teide~1 (Rebolo et al. 1996), and a factor 2.5 
lower than the EW measured in PPl~15 (Basri et al. 1996). 
Thus, we are confident that CFHT-Pl~18 does not pass the lithium test.  
This binary system appears to be located at precisely 
the right distance to be confused with a Pleiades member. We do not know the distance 
to this system because it is not a cluster member, but we can estimate it from 
the spectral type (M8, Mart\'\i n et al. 1998a). If it has the same absolute J-band 
magnitude as LHS~2397a (M8, M$_{\rm J}$=11.13, Leggett et al. 1998), 
we obtain a spectroscopic parallax of 105~pc. The projected separation of the binary 
would then be 34.5~AU. 

CFHT-Pl~16 is located more than 0.5 mag above the Pleiades sequence in all the CMDs. 
It could be  an unresolved binary with nearly identical components. 
Alternatively, CFHT-Pl~16 could be a non-member. 
We obtained Keck/LRIS mid-resolution spectroscopy to test its membership. 
The reduced spectrum is shown 
in Fig.~5. We clearly detect H$_\alpha$ in emission with EW=-6.1~A. 
The quality of our spectrum is barely  
sufficient for a detection of the LiI resonance line. There is an absorption 
feature at the position of the LiI resonance line with EW$\sim$1.2~A, which is similar to the EW of 
Teide~1 (Rebolo et al. 1996), but there are also noise 
features of similar strength elsewhere in the spectrum. We estimate that the probability of having 
a noise feature at the position of the LiI line is $<$10\%, i.e. low but not negligible. 
We measured a heliocentric radial velocity 
of 13$\pm$5 km/s using the H$_\alpha$ feature, which is consistent with the radial 
velocities of VLM Pleiades members (Mart\'\i n et al. 1998b).

\subsection{Search for Companions} 

Since the NICMOS PSF can vary significantly from one HST orbit to the next,
we used all of the HST/NICMOS pointings to build up a library of
HST/NIC1 PSFs in all three filters (F110M, F145M, anf F165M). For
each set of observation, we then identified the best matching
PSF in the library for PSF subtraction. Sub-pixel offsets between
target and PSF were computed by cross-correlating the individual frames.
The PSF was then Fourier-shifted, scaled, and subtracted.

Fig.~6 shows examples of NIC1 images before and after 
point spread function (PSF) subtraction. 
The dark line seen in each image is the boundary between the individual NICMOS
quadrants. In the original image we can  
clearly see the first and second airy rings. They are considerably 
reduced in the subtracted image.  

We have estimated limiting sensitivities to the 
presence of companions as a function of separation from 
the primary. They are given in Table~5 for two different separations. 
A few radial averages of the 
limits to the presence of companions in NICMOS images are shown in Fig.~7. 
They were derived under the assumption
that we would detect a source if the counts in the central pixel of its PSF
are at least 3 times the sigma of the background. 
The four objects, namely CFHT-Pl 15, CFHT-Pl 19, 
HHJ 2 and MHO 3,  for which the residuals of the PSF subtraction were 
high have been flagged in Table~5. The residuals could be due to the presence of faint companions 
within 5 pixels (0.215") of the primary, or to unrepeatable structure in the PSF. 

We have not found any distant (sep.$>$1") companions in the near-IR images 
taken at CFHT and Mt Hopkins. Any companion up to 3 magnitudes fainter than the primary 
in the separation range 1 to 8 arcseconds, should have been detected. 

\section{Membership in the Pleiades}

\subsection{Membership criteria}

The membership criteria that we have used are the following: 
1) Is the object a dwarf or a giant? All the objects 
for which we have low-resolution spectra have KI and NaI lines similar to those  
of dwarfs. Because of the low-resolution of most of our spectra, 
the NaI doublet at 818.3 and 819.0~nm 
is blended with telluric absorption and we cannot derive an accurate equivalent width. 
Thus, we do not use the strength of the line as a membership criterion 
(Mart\'\i n et al. 1996). 
2) Is the radial velocity 
similar to those of known members? We have used radial velocity data from the literature 
(Basri et al. 1996; Mart\'\i n et al. 1998b; Stauffer et al. 1998b). 3) 
Is the proper motion similar to the cluster? We have used proper motion data from the 
literature (Hambly et al. 1993; Hambly et al. 1999). 4) Does the object have lithium?
Pleiades members fainter than I$\sim$17.8 should have lithium 
(Mart\'\i n et al. 1998b; Stauffer et al. 1998b). 
Brighter objects could have lithium if they are binaries (PPl~1). 
5) Does the object have H$_\alpha$ in emission? The majority of Pleiades 
members do show H$_\alpha$ in emission because of their young age. However, H$_\alpha$ is 
variable and sometimes not detected. For example, we did not detect it in our Keck spectrum 
of Teide~1, but it was previously detected by Rebolo et al. (1996). We do not reject any 
candidate member solely on basis of lack of H$_\alpha$ emission, but we note that many 
non-members do not have H$_\alpha$ emission, suggesting that they are relatively old 
field dwarfs. 6) We checked the position of each object in the spectral type versus apparent 
I-band magnitude (Fig.~4). The lithium BDs are considered as benchmark objects, and most 
of the BD candidates are located near them. 
Objects that deviate from the sequence by more than 0.8 magnitudes 
are considered as likely non-members. 7) The location of the lithium BDs in the 
I versus (I-K) and F110M versus (F110M-F165M) CMDs is well fitted with the Nextgen isochrone. 
On the other hand, the location of Roque~25 in the same CMDs is fitted by the Dusty isochrone. 
Objects that are 3$\sigma$ away from the isochrones are considered as likely non-members. 

We summarize our assessment of the membership status of 45 objects in Table~6. 
We confirm the Pleiades membership of 37 VLM stars and BDs, and reject 8 objects as non-members. The 
later objects are excluded from our calculation of binary statistics in the Pleiades cluster.

\subsection{Physical Parameters}

In order to derive the physical parameters of the objects that we consider 
as likely Pleiades members, we adopted a common 
age, distance and metallicity of 120~Myr, 125~pc and solar, respectively. 

The only way to infer masses for our objects is to compare their location in the CMD 
with theoretical calculations. In Fig.~8,
 we show the theoretical mass-luminosity relationships for several near-IR filters 
that we have adopted. 
The behaviour of the NICMOS filters F110M and F165M mimics that of the $J$ 
and $H$ broad-band filters, respectively. 
The sensitivity to dust effects changes with $\lambda$ and mass (temperature) 
across these plots. We have  
minimized the dependence of our results on the role of dust in the atmosphere by using 
the filters that show less sensitivity to the choice of model (Nextgen or Dusty) for a given mass. 
For all the objects we used the absolute magnitudes in the J and 
F110M filters (we averaged the two of them if both magnitudes were available) and 
the NEXTGEN models, but for Roque~25 we used the K-band absolute 
magnitude and the DUSTY model.  
Our mass estimates for the CFHT objects are consistent within $\pm$0.005~M$_\odot$ with those 
derived by B98 using three different sets of models and absolute I-band magnitudes. 

\section{Binary statistics}

The masses of the primaries considered in this study range from 0.14~M$_\odot$ 
(CFHT-Pl-1, I=16.1, B98) to 0.035~M$_\odot$ (Roque~25, I=21.17, Mart\'\i n et al. 1998c). 
The masses of the primaries observed with high-resolution techniques (AO and HST) 
are all lower than 0.11~M$_\odot$ (HHJ19, I=16.7, Hambly et al. 1993). 
We estimated upper limits to the masses of undetected companions using the 
DUSTY models and the sensitivity limits in the $K$ and/or F165M filters. 
The average limits on the mass ratios (q=M$_2$/M$_1$) for different separation ranges  
are given in Table~7. The limits on close ($<$1") companions come from the HST and AO images, and 
the limits on distant companions were derived from the ground-based K-band images. 

Since the only binary that we found, CFHT-Pl-18, did not pass the lithium test, 
we have found zero resolved binaries in our sample. 
However, this null result does not necessarily imply that the binary frequency among 
Pleiades VLM members is very low. Basri \& Mart\'\i n (1999) found that PPl~15 is 
a double-lined binary with a period of only 5.8 days. Zapatero Osorio et al. (1997a) 
had noticed that this object lies on the binary sequence. Five other Pleiades VLM members  
also lie in the binary sequence in the CMD of 
Fig.~2. We consider them as likely 
binaries with nearly-identical components. Two of them (lying near HHJ~6) 
have not been observed with HST or AO. 
Three of them (HHJ~6, CFHT-Pl-12 and 16) are not resolved 
with HST or CFHT/AO. Steele \& Jameson (1995) have suggested 
that 46\% of Pleiades VLM stars are multiple on the basis of a spectroscopic study of 
temperature indicators. However,  none of their  
HHJ candidate binaries observed with HST/WFPC were resolved.

Bouvier, Rigaut \& Nadeau  (1997) carried out an AO search 
for binaries among 144 
G- and K-type Pleiades members. They found that the binary frequency and period distribution 
of their sample is similar to that of DM91. Since the DM91 study has better statistics, 
we have used their orbital period and mass ratio distributions 
for predicting the number of binaries expected in our sample. 
The results are given in the third column of Table~7. We have converted the observed upper limits 
on components separations on the plane of the sky (s) to semi-major axes (a) using a cluster distance 
of 125~pc and s/a=0.92 (Heacox \& Gathright 1994). 
Considering our sample of 
Pleiades BDs observed with HST/NICMOS, we find that there is a factor of $\sim$3 excess of 
binaries with respect to the distribution of DM91 for separations smaller than 10.9~AU, 
which corresponds to orbital periods shorter than $\sim$110 years.  
However, because of the small size of our sample we find that the difference between the Pleiades 
and the field binaries is significant only at the 1~$\sigma$ level. 
On the other hand, it is more significant that we do not find any binaries with separations 
larger than $\sim$27~AU. The lack of Pleiades VLM systems with periods longer than 444 years 
(3.4 expected, 0 found) is a 2~$\sigma$ effect for Poisson statistics. 

Reid \& Gizis (1997) obtained HST observations of 53 Hyades M-type members (primary mass $\le$0.3~M$_\odot$). 
They found nine binaries with q$>$0.5 in the separation range 14 to 825~AU, 
consistent with observations of M-dwarfs in the solar neighborhood.  
If the binary properties of our Pleiades VLM sample was similar to 
that of the Hyades sample of Reid \& Gizis, we should have found 2 binaries with separations larger 
than 30~AU. 

When they grow old, 
the Pleiades VLM stars and BDs will become cooler and will look similar to L-type 
field dwarfs. The first three field 
binaries with L-dwarf components have been recently discovered (Mart\'\i n, Brandner \& 
Basri 1999; Koerner et al. 1999). All of them have separations less than 10 AU, and would 
not have been resolved in our survey if they were at the Pleiades distance.   
There is no bias against finding wider binaries because 
the FOV in these searches is $\ge$100~AU. 
In fact, no wide doubles have been reported in the infrared images of $\sim$ 100 L dwarfs 
and $\sim$15 methane dwarfs identified so far in the DENIS, Sloan and 2MASS surveys. 
However, there has not been a survey for spectroscopic binaries among ultracool dwarfs. 
There could be many short-period VLM binaries that have not been identified yet. 
The null result of our survey for Pleiades BD binaries is consistent with the properties of the 
field BD binaries found so far.  
    
The median orbital separation of the main-sequence stars studied by DM91 is 23~AU (Heacox 1998). 
Two Pleiades BD binaries (PPl~15 and CFHT~Pl~16) and three field BD binaries have been 
identified with separations $<$23~AU, but no BD binaries are known to have separations $\ge$23~AU. 
The systematically small separations of the BD binaries with respect to the DM91 sample, 
suggest that there is a difference in population characteristics. The small numbers involved preclude 
any definitive conclusion in this regard, but it is worth of mention and brief discussion. 
 
The mass of the 
primary could be an important factor in the the process of binary formation. 
Basri \& Mart\'\i n (1999) suggested that the formation of substellar binaries is biased 
toward smaller separations than the formation of stellar binaries because the probability 
of finding a short-period binary like PPl~15 from a normal (DM91-type) binary distribution 
is only $\sim$10\%.  Another interesting 
proposition is that of a ``brown dwarf desert'', i.e. a scarcity of BD secondaries 
to G and K dwarfs within 2 AU of the primary (Marcy \& Butler 1998). It seems that BDs, as objects 
representative of the tail of the stellar mass function, could provide important hints about 
binary formation. 

We briefly consider three possible scenarios that could explain a difference in the characteristics 
of BD binaries with respect to the DM91 sample:  
(1) Disruption of wide binaries 
due to dynamical interactions with the stellar members of the cluster. This process has been 
invoked by Kroupa, Petr \& McCaughrean (1999) to explain the binary frequency of stars in the 
Trapezium cluster. An important test of this scenario is the binary frequency of hard binaries, 
which should be independent of primary mass. (2) If 
VLM stars and BDs usually come from the dynamical ejection of VLM fragments from protostellar aggregates 
(Laughlin \& Lin, private communication), the ones that are more likely to survive are close 
binaries with nearly-identical component. Such an scenario has two constraints to 
overcome. The typical ejection velocity of the ejected VLM star or BD cannot be much larger than the escape 
velocity from the Pleiades cluster. The mechanism has to be extremely efficient because BDs are 
quite numerous. (3) Numerical simulations of molecular cloud fragmentation 
have shown that slower rotating cloud cores form closer binaries than when 
the clouds are rapidly rotating (Boss 1993). Small cloud cores that start collapse from 
high initial densities form even closer binaries. This scenario seems provide a natural way 
of explaining the absence of wide BD binaries and an excess of close systems. Furthermore, 
the BD desert around solar-type stars can also be explained by fragmentation processes (Bate 2000).

\section{The Pleiades substellar mass function}

We have presented NIR photometry and low-resolution spectroscopy for all the BD candidates 
reported by B98. We have also applied the lithium test to CFHT-Pl-16 and 18. 
The membership status of all the CFHT BD candidates is summarized in Table~6. 
We restrict ourselves to the CFHT sample for addressing the issue of the substellar mass function, 
because the level of follow-up observations 
of the BD candidates coming from other surveys is still incomplete. 

The level of field star contamination among Pleiades BD candidates 
was assessed by B98 using two statistical approaches. They compared 
the mass of Pleiades members and field stars in the same volume, and they used a field luminosity 
function. Both methods were in good agreement, and suggested a contamination of 25\% . 
We confirm the membership of 12 Pleiades CFHT BD candidates out of the 18 
included in B98. The 6 non-members are all foreground cool M dwarfs. We find a contamination of 
33\%, which is somewhat higher than the estimate of B98. 
These authors obtained a power-law index in a log-log plot (dN/dM$\sim$M$^{\alpha}$)
of $\alpha$=-0.6 for the cluster IMF across the substellar boundary. 
We revise this slope downwards to a value of $\alpha$=-0.53 , implying that the number of BDs per 
decreasing mass bin is moderately rising, but their relative contribution to the total mass of the Pleiades 
cluster is diminishing. 

We cannot make a detailed correction of the IMF for the binary fraction of the CFHT BDs because we have not 
resolved any system. Nevertheless, we believe that the binary corrections are probably not very important 
because there does not seem to be many BD companions to stars, and because 
our search for binaries suggest that BDs do not have a higher 
multiplicity frequency than M-type stars.

\acknowledgments

{\it Acknowledgments}: 
This research makes use of data collected at the 
Hubble Space Telescope, the Canada-France-Hawaii telescope, the Lick Nickel 
telescope and the Mt Hopkins telescope. 
Some of the data presented herein were
obtained at the W.M. Keck Observatory, which is operated as a scientific
partnership among the California Institute of Technology, the University of
California and the National Aeronautics and Space Administration.  The
Observatory was made possible by the generous financial support of the W.M.
Keck Foundation.
We are grateful to the HST staff  (particularly, Beth Perriello 
and Al Schultz) for their assistance. We thank 
France Allard, Isabelle Baraffe, Gilles Chabrier and Peter Hauschildt  
for making their evolutionary models available to us. 
EM thanks Alan Boss, Greg Laughlin and Doug Lin for discussions about the formation of 
brown dwarf binaries.  
GB acknowledges the support of NSF through grant AST96-18439. 
FA was supported by NASA through grants 110-96LTSA and NAG5-3435.
Funding for this publication was provided by NASA through Proposal 
GO-7899 submitted to the Space Telescope Science Institute, which is operated 
by the Association of UNiversities for Research in Astronomy, Inc., under
NASA contract NAS5-26555. 


\begin{deluxetable}{lcccc}
\footnotesize
\tablecaption{\label{tab1} NICMOS photometry for Pleiades VLM candidates}
\tablewidth{0pt}
\tablehead{
\colhead{Name}  &
\colhead{Other Name}  &
\colhead{m(F110M)} &
\colhead{m(F145M)}  &
\colhead{m(F165M)}        }                            
\startdata 
Calar 3\tablenotemark{*} & CFHT 21  & 17.07$\pm$0.16 & 16.76$\pm$0.40 & 16.24$\pm$0.20 \\ 
CFHT-Pl 15      &          & 16.53$\pm$0.01 & 15.87$\pm$0.01 & 15.22$\pm$0.01 \\
CFHT-Pl 16      &          & 16.28$\pm$0.04 & 15.61$\pm$0.02 & 14.95$\pm$0.02 \\
CFHT-Pl 17      &          & 16.76$\pm$0.05 & 16.03$\pm$0.01 & 15.43$\pm$0.01 \\ 
CFHT-Pl 18 A    &          & 16.96$\pm$0.07 & 16.31$\pm$0.03 & 15.77$\pm$0.04 \\
CFHT-Pl 18 B    &          & 17.74$\pm$0.14 & 17.11$\pm$0.07 & 16.50$\pm$0.07 \\
CFHT-Pl 19      &          & 16.93$\pm$0.09 & 16.37$\pm$0.01 & 15.77$\pm$0.02 \\
CFHT-Pl 20      &          & 17.17$\pm$0.05 & 16.40$\pm$0.09 & 15.90$\pm$0.02 \\
CFHT-Pl 22      &          & 17.41$\pm$0.09 & 17.03$\pm$0.03 & 16.53$\pm$0.02 \\
CFHT-Pl 23      &          & 17.09$\pm$0.05 & 16.31$\pm$0.09 & 15.71$\pm$0.02 \\
CFHT-Pl 25      &          & 17.30$\pm$0.05 & 16.59$\pm$0.03 & 15.93$\pm$0.03 \\
HHJ  2       &          & 15.85$\pm$0.04 & 15.19$\pm$0.01 & 14.75$\pm$0.01 \\
HHJ  8       &          & 15.53$\pm$0.02 & 14.86$\pm$0.01 & 14.42$\pm$0.01 \\
MHO  1       &          & 15.90$\pm$0.06 & 15.41$\pm$0.01 & 14.91$\pm$0.01 \\
MHO  3       &          & 16.08$\pm$0.08 & 15.51$\pm$0.02 & 14.90$\pm$0.01 \\
MHO  6       &          & 15.97$\pm$0.04 & 15.32$\pm$0.02 & 14.83$\pm$0.02 \\
PIZ 1        &          & 17.11$\pm$0.11 & 16.53$\pm$0.13 & 15.79$\pm$0.05 \\
PPl 1        & Roque 15 & 15.97$\pm$0.01 & 15.48$\pm$0.01 & 14.73$\pm$0.01 \\
             &          & 15.90$\pm$0.03 & 15.31$\pm$0.01 & 14.71$\pm$0.01 \\
PPL 14       &          & 15.82$\pm$0.03 & 15.33$\pm$0.01 & 14.82$\pm$0.01 \\	
Roque 4      &          & 17.24$\pm$0.04 & 16.55$\pm$0.04 & 15.86$\pm$0.02 \\
Roque 7    & CFHT-Pl 24 & 17.12$\pm$0.03 & 16.43$\pm$0.02 & 15.79$\pm$0.01 \\
Roque 11     &          & 16.65$\pm$0.03 & 16.00$\pm$0.02 & 15.43$\pm$0.02 \\
Roque 12     &          & 16.59$\pm$0.05 & 15.89$\pm$0.03 & 15.34$\pm$0.02 \\
Roque 13     &          & 16.27$\pm$0.01 & 15.67$\pm$0.01 & 15.04$\pm$0.01 \\
Roque 14     &          & 16.11$\pm$0.05 & 15.50$\pm$0.07 & 14.87$\pm$0.04 \\
Roque 16   & CFHT-Pl 11 & 16.16$\pm$0.03 & 15.49$\pm$0.03 & 14.97$\pm$0.01 \\
Roque 17     &          & 15.91$\pm$0.04 & 15.31$\pm$0.01 & 14.73$\pm$0.01 \\
Roque 25     &          & 18.32$\pm$0.23 & 17.77$\pm$0.09 & 16.77$\pm$0.02 \\
Teide 1      &          & 16.92$\pm$0.03 & 16.18$\pm$0.01 & 15.59$\pm$0.01 \\
Teide 2    & CFHT-Pl 13 & 16.18$\pm$0.04 & 15.48$\pm$0.01 & 14.97$\pm$0.01 \\ 
\enddata
\tablenotetext{*}{Trailed image because of guiding error} 
\tablecomments{The error bars are 1~$\sigma$ standard deviations.}
\end{deluxetable}

\begin{deluxetable}{lcccc}
\footnotesize
\tablecaption{\label{tab2} New ground-based photometry for Pleiades VLM candidates}
\tablewidth{0pt}
\tablehead{
\colhead{Name}  &
\colhead{m($J$)} &
\colhead{m($H$)}  &
\colhead{m($K$)} &
\colhead{Telescope}       }                            
\startdata 
Calar 3      & 16.10$\pm$0.02 & 15.42$\pm$0.04 & 14.90$\pm$0.07 & CFHT \\
CFHT-Pl  1      & 14.41$\pm$0.03 &                & 13.50$\pm$0.10 & MHO  \\
CFHT-Pl  2      & 14.69$\pm$0.03 &                & 13.71$\pm$0.10 & MHO  \\
CFHT-Pl  3      & 14.81$\pm$0.03 &                & 13.82$\pm$0.10 & MHO  \\
CFHT-Pl  4      & 14.96$\pm$0.03 &                & 14.01$\pm$0.10 & MHO  \\
CFHT-Pl  5      & 15.08$\pm$0.03 &                & 14.06$\pm$0.10 & MHO  \\
CFHT-Pl  6      & 14.89$\pm$0.03 &                & 13.78$\pm$0.10 & MHO  \\
CFHT-Pl  7      & 15.42$\pm$0.03 &                & 14.43$\pm$0.10 & MHO  \\
CFHT-Pl  8      & 15.16$\pm$0.03 &                & 14.13$\pm$0.10 & MHO  \\
CFHT-Pl  9      & 15.46$\pm$0.10 &                &                & Lick \\
             & 15.44$\pm$0.03 &                & 14.44$\pm$0.10 & MHO  \\
CFHT-Pl 10      & 15.49$\pm$0.10 &                &                & Lick \\
             & 15.53$\pm$0.05 &                & 14.49$\pm$0.10 & MHO  \\
CFHT-Pl 12      & 15.10$\pm$0.02 & 14.51$\pm$0.04 & 14.11$\pm$0.07 & CFHT \\
             & 15.17$\pm$0.10 &                &                & Lick \\
             & 15.23$\pm$0.03 &                & 14.14$\pm$0.10 & MHO  \\
CFHT-Pl 15      & 16.00$\pm$0.02 & 15.22$\pm$0.04 & 14.87$\pm$0.07 & CFHT \\
             & 15.98$\pm$0.03 &                & 14.80$\pm$0.10 & MHO  \\
CFHT-Pl 16      & 15.65$\pm$0.02 & 14.85$\pm$0.04 & 14.39$\pm$0.07 & CFHT \\
             & 15.69$\pm$0.10 &                &                & Lick \\ 
CFHT-Pl 17      & 15.98$\pm$0.02 & 15.20$\pm$0.04 & 14.90$\pm$0.07 & CFHT \\
             & 16.05$\pm$0.10 &                &                & Lick \\
CFHT-Pl 18      & 15.95$\pm$0.02 & 15.23$\pm$0.04 & 14.80$\pm$0.07 & CFHT \\          
CFHT-Pl 19      & 16.51$\pm$0.02 & 15.70$\pm$0.04 & 15.47$\pm$0.07 & CFHT \\
             & 16.48$\pm$0.10 &                &                & Lick \\
CFHT-Pl 20      & 16.56$\pm$0.02 & 15.88$\pm$0.04 & 15.56$\pm$0.07 & CFHT \\
             & 16.54$\pm$0.10 &                &                & Lick \\
CFHT-Pl 22      & 17.11$\pm$0.02 & 16.61$\pm$0.03 & 15.95$\pm$0.10 & CFHT \\
CFHT-Pl 23      & 16.46$\pm$0.02 & 15.63$\pm$0.04 & 15.24$\pm$0.07 & CFHT \\
CFHT-Pl 25      & 16.68$\pm$0.02 & 15.87$\pm$0.04 & 15.48$\pm$0.07 & CFHT \\
Roque 16     & 15.52$\pm$0.02 & 14.87$\pm$0.04 & 14.50$\pm$0.07 & CFHT \\
             & 15.67$\pm$0.03 &                & 14.56$\pm$0.10 & MHO  \\
\enddata
\tablecomments{The error bars are 1~$\sigma$ standard deviations.}
\end{deluxetable}

\begin{deluxetable}{lcccccc}
\footnotesize
\tablecaption{\label{tab3} Spectroscopic log} 
\tablewidth{0pt}
\tablehead{
\colhead{Name}  &
\colhead{Date}  & 
\colhead{Telescope} &
\colhead{Grating}  &
\colhead{Disp. (\AA /pix)}  & 
\colhead{Range (nm)}  &
\colhead{$\tau_{\rm{exp}}$ (s)}
                              }
\startdata
CFHT-Pl 1       & 1998 Dec 22  & KPNO    & 730  & 4.20 & 533.9--959.7  & 900 \\
CFHT-Pl 2       & 1998 Dec 22  & KPNO    & 730  & 4.20 & 533.9--959.7  & 900 \\
CFHT-Pl 5       & 1998 Dec 23  & KPNO    & 730  & 4.20 & 533.9--959.7  & 900 \\
CFHT-Pl 6       & 1998 Dec 23  & KPNO    & 730  & 4.20 & 533.9--959.7  & 900 \\
CFHT-Pl 7       & 1998 Dec 22  & KPNO    & 730  & 4.20 & 533.9--959.7  & 1800 \\
CFHT-Pl 8       & 1998 Dec 22  & KPNO    & 730  & 4.20 & 533.9--959.7  & 1200 \\
CFHT-Pl 16      & 1998 Dec 22  & KPNO    & 730  & 4.20 & 533.9--959.7  & 1800 \\
CFHT-Pl 16      & 2000 Jan 5   & Keck II & 900  & 0.85 & 631.2--802.9  & 3600 \\
CFHT-Pl 17      & 1998 Dec 22  & KPNO    & 730  & 4.20 & 533.9--959.7  & 1800 \\
CFHT-Pl 18      & 1998 Dec 21  & Keck II & 900  & 0.85 & 631.2--802.9  & 5400 \\
CFHT-Pl 19      & 1998 Dec 22  & KPNO    & 730  & 4.20 & 533.9--959.7  & 1800 \\
CFHT-Pl 20      & 1998 Dec 23  & KPNO    & 730  & 4.20 & 533.9--959.7  & 1800 \\
CFHT-Pl 25      & 1998 Dec 23  & Keck II & 150  & 4.82 & 355.5--1125.1 & 1200 \\
CFHT-Pl 26      & 1998 Dec 23  & Keck II & 150  & 4.82 & 355.5--1125.1 & 1800 \\
Roque 7      & 1997 Dec 28  & WHT     & 158  & 2.91 & 629.1--926.5 & 2000 \\
Roque 25     & 1998 Dec 23  & Keck II & 150  & 4.82 & 355.5--1125.1 & 1800 \\
Roque 33     & 1998 Dec 23  & Keck II & 150  & 4.82 & 355.5--1125.1 & 2000 \\
Teide 1      & 1998 Dec 23  & Keck II & 150  & 4.82 & 355.5--1125.1 & 1200 \\
\enddata
\end{deluxetable}

\begin{deluxetable}{lccccc}
\footnotesize
\tablecaption{\label{tab4} Spectroscopic data} 
\tablewidth{0pt}
\tablehead{
\colhead{Name}  &
\colhead{PC3}  & 
\colhead{SpT} &
\colhead{TiO}  &
\colhead{VO}   & 
\colhead{H$_\alpha$}
                              }
\startdata
              &       &        &      &      & (\AA )         \\
CFHT-Pl 1        & 1.26  &  dM4.9 & 2.56 & 2.41 & -3.1$\pm$0.5   \\
CFHT-Pl 2        & 1.26  &  dM4.9 & 3.09 & 2.43 & -3.4$\pm$0.5   \\
CFHT-Pl 5        & 1.36  &  dM5.5 & 3.26 & 2.52 & -4.0$\pm$0.5   \\
CFHT-Pl 6        & 1.61  &  dM6.9 & 3.52 & 2.70 &  $\ge$-2       \\
CFHT-Pl 7        & 1.37  &  dM5.6 & 3.08 & 2.43 &  $\ge$-2       \\
CFHT-Pl 8        & 1.37  &  dM5.6 & 3.09 & 2.51 & -14.6$\pm$0.4  \\
CFHT-Pl 16       & 2.20  &  dM9.3 & 4.41 & 2.74 & $\ge$-15       \\
CFHT-Pl 16\tablenotemark{*}        &       &        &      &      & -6.1$\pm$0.3   \\
CFHT-Pl 17       & 1.82  &  dM7.9 & 4.48 & 2.86 & -7:            \\
CFHT-Pl 18\tablenotemark{*}       &       &        &      &      & -3.0$\pm$0.2   \\
CFHT-Pl 19       & 1.68  &  dM7.5 & 4.07 & 2.50 &  $\ge$-10      \\
CFHT-Pl 20       & 1.71  &  dM7.5 & 3.55 & 2.51 &  $\ge$-3       \\
CFHT-Pl 25       & 2.10  &  dM9.0 & 4.75 & 3.06 &  $\ge$-6       \\
CFHT-Pl 26       & 1.65  &  dM7.1 & 3.76 & 3.20 &  $\ge$-5       \\
Roque 7       & 1.90  &  dM8.3 & 4.60 & 2.89 & $\ge$-15       \\
Roque 25      & 2.58  &  dL0.1 & 2.61 & 2.51 &  -9$\pm$2      \\
Roque 33      & 2.44  &  dM9.8 & 6.35 & 3.20 &  -55$\pm$10    \\
Teide 1       & 1.81  &  dM7.9 & 4.22 & 2.80 &  -2:           \\
\enddata
\tablenotetext{*}{Measurement using intermediate resolution spectrum} 
\tablecomments{A colon indicates uncertain detection due to low S/N ratio or blending.}
\end{deluxetable}

\begin{deluxetable}{lcccc}
\footnotesize
\tablecaption{\label{tab5} Limits to the presence of companions in the NIC1 frames}
\tablewidth{0pt}
\tablehead{
\colhead{Name}  &
\colhead{$\Delta$ m(F110M) } &
\colhead{$\Delta$ m(F165M) } &
\colhead{$\Delta$ m(F110M) } &
\colhead{$\Delta$ m(F165M) }       }                            
\startdata 
                           & (0".1) & (0".1)  & (0".4) & (0".4) \\
Calar 3                    &     &     & 1.0 & 1.0 \\ 
CFHT-Pl 15\tablenotemark{*}   & 2.6 & 3.4 & 5.2 & 4.5 \\
CFHT-Pl 16                    & 4.3 & 4.5 & 4.8 & 5.0 \\
CFHT-Pl 17                    & 4.0 & 4.6 & 4.9 & 5.6 \\
CFHT-Pl 19\tablenotemark{*}   & 2.1 & 3.7 & 4.6 & 4.4 \\
CFHT-Pl 22                    & 3.5 & 4.0 & 4.2 & 4.2 \\
CFHT-Pl 23                    & 3.9 & 4.1 & 4.7 & 5.2 \\
CFHT-Pl 25                    & 4.2 & 5.1 & 5.2 & 5.5 \\
HHJ 2\tablenotemark{*}     & 3.3 & 4.5 & 5.6 & 5.5 \\
HHJ 8                      & 3.0 & 3.0 & 4.5 & 4.5 \\
MHO 1                      & 3.7 & 4.7 & 5.0 & 5.3 \\
MHO 3\tablenotemark{*}     & 2.9 & 3.8 & 5.1 & 5.5 \\
MHO 6                      & 3.0 & 3.0 & 4.5 & 4.5 \\
PIZ 1                      & 3.1 & 4.1 & 3.5 & 4.4 \\
PPl 1                      & 4.6 & 4.8 & 5.8 & 5.8 \\
PPl 14                     & 4.6 & 4.7 & 5.2 & 5.4 \\
Roque 4                    & 4.2 & 4.4 & 4.7 & 5.2 \\
Roque 7                    & 3.9 & 4.1 & 4.7 & 5.5 \\
Roque 11                   & 3.0 & 3.0 & 4.5 & 4.5 \\
Roque 12                   & 3.0 & 3.0 & 4.5 & 4.5 \\
Roque 13                   & 4.2 & 4.6 & 5.4 & 5.5 \\
Roque 14                   & 4.1 & 4.5 & 5.3 & 5.7 \\
Roque 16                   & 4.6 & 4.4 & 5.7 & 5.6 \\
Roque 17                   & 3.9 & 4.2 & 5.5 & 5.2 \\
Roque 25                   & 3.0 & 3.8 & 3.3 & 3.9 \\
Teide 1                    & 4.1 & 4.1 & 5.4 & 4.9 \\
Teide 2                    & 4.0 & 4.3 & 4.7 & 5.2 \\
\enddata
\tablenotetext{*}{Residuals in the PSF subtraction, possibly due to the presence of an 
unresolved companion with brightness close to   
the $\Delta m$ limit given in this Table and separation less than 0".22}
\end{deluxetable}

\begin{deluxetable}{lccccccccc}
\footnotesize
\tablecaption{\label{tab6} Membership status}
\tablewidth{0pt}
\tablehead{
\colhead{Name}  &
\colhead{Dwarf?} &
\colhead{V$_{rad}$} &
\colhead{pm} &
\colhead{Li} &
\colhead{H$_\alpha$} &
\colhead{SpT} &
\colhead{I-K} &
\colhead{NIC}  &
\colhead{Member?}       
}                            
\startdata 
Calar 3       & Yes & Yes &     & Yes & Yes  & Yes &  Yes &        & Yes \\
CFHT-Pl  1       & Yes &     &     &     & Yes  & Yes &  Yes &     & Yes \\
CFHT-Pl  2       & Yes &     &     &     & Yes  & Yes &  Yes &     & Yes \\
CFHT-Pl  3\tablenotemark{*}       & Yes &     & Yes &     &      &     &  Yes &     & Yes \\
CFHT-Pl  4       &     &     &     &     &      &     &  Yes &     & Yes? \\
CFHT-Pl  5       & Yes &     &     &     & Yes  & Yes &  Yes &     & Yes \\
CFHT-Pl  6       & Yes &     &     &     & No   & Yes &  Yes &     & Yes? \\
CFHT-Pl  7       & Yes &     &     &     & No   & Yes &  Yes &     & Yes? \\
CFHT-Pl  8       & Yes &     &     &     & Yes  & Yes &  Yes &     & Yes \\
CFHT-Pl  9       & Yes &     &     & No  & Yes  &     &  Yes &     & Yes \\
CFHT-Pl 10       & Yes &     &     & No  & Yes  &     &  Yes &     & Yes \\
CFHT-Pl 12       & Yes &     &     & Yes & Yes  &     &  Yes & Yes & Yes \\
CFHT-Pl 14       & Yes &     & No  & No  & No   &     &      &     & No  \\
CFHT-Pl 15       & Yes &     &     & Yes & Yes  &     &  Yes & Yes & Yes \\
CFHT-Pl 16       & Yes & Yes &     & Yes & Yes  & Yes &  Yes & Yes & Yes  \\
CFHT-Pl 17       & Yes &     &     &     & Yes  & Yes &  Yes & Yes & Yes \\
CFHT-Pl 18       & Yes &     &     & No  & Yes  & Yes &  Yes & No  & No \\
CFHT-Pl 19       & Yes &     &     &     &      & Yes &  No  & Yes & No? \\
CFHT-Pl 20       & Yes &     &     &     & No   & Yes &  No  & No  & No \\
CFHT-Pl 22       & Yes &     &     &     &      & Yes &  No  & No  & No \\
CFHT-Pl 23       & Yes &     &     &     &      &     &  Yes & Yes & Yes \\
CFHT-Pl 25       & Yes &     &     &     & No   & Yes &  Yes & Yes & Yes \\
CFHT-Pl 26       & Yes &     &     &     & No   & No  &      &     & No  \\
HHJ 2         & Yes &     & Yes &     & Yes  & Yes &  Yes & Yes & Yes \\
HHJ 6         & Yes & Yes & Yes &     & Yes  & Yes &  Yes & Yes & Yes \\
HHJ 8         & Yes &     & Yes &     & Yes  & Yes &  Yes & Yes & Yes \\
MHO 1         &     &     &     &     &      &     &      & No  & No  \\
MHO 3         & Yes & Yes &     & Yes & Yes  &     &      & Yes & Yes \\
MHO 6         &     &     &     &     &      &     &      & Yes & Yes? \\
PIZ 1         & Yes &     &     &     &      & Yes &  Yes & Yes & Yes \\
PPl 1         & Yes & Yes &     & Yes & Yes  & Yes &  Yes & Yes & Yes \\
PPl 14        & Yes &     &     &     & Yes  & Yes &      & No  & No? \\
PPl 15        & Yes & Yes & Yes & Yes & Yes  & Yes &  Yes &     & Yes \\
Roque  4      & Yes & Yes &     &     & Yes  & Yes &  Yes & Yes & Yes \\
Roque  7      & Yes &     &     &     &      & Yes &      & Yes & Yes \\
Roque 11      & Yes & Yes &     &     & Yes  & Yes &  Yes & Yes & Yes \\
Roque 12      & Yes &     &     &     & Yes  & Yes &  Yes & Yes & Yes \\
Roque 13      & Yes &     &     & Yes & Yes  &     &  Yes & Yes & Yes \\
Roque 14      & Yes &     &     & Yes & Yes  & Yes &      & Yes & Yes \\
Roque 16      & Yes &     &     & Yes & Yes  &     &  Yes & Yes & Yes \\
Roque 17      & Yes &     &     & Yes & Yes  & Yes &      & Yes & Yes \\
Roque 25      & Yes &     &     &     & Yes  & Yes &  Yes & Yes & Yes \\
Roque 33      & Yes &     &     & Yes & Yes  & Yes &      &     & Yes \\
Teide 1       & Yes & Yes & Yes & Yes & Yes  & Yes &  Yes & Yes & Yes \\
Teide 2       & Yes & Yes &     & Yes & Yes  & Yes &  Yes & Yes & Yes \\
\enddata
\tablenotetext{*}{Dwarf status inferred from proper motion because a spectrum is 
not available.}
\end{deluxetable}

\begin{deluxetable}{lcccccc}
\footnotesize
\tablecaption{\label{tab7} Binary frequency}
\tablewidth{0pt}
\tablehead{
\colhead{N$_{\rm objects}$}  &
\colhead{N$_{\rm binaries}$}  &
\colhead{N$_{\rm expected}$}  &
\colhead{Sep. range (")}  &
\colhead{Dist. range (AU)}  &
\colhead{q$_{\rm min}$} &
\colhead{log P (d)}  
        }                            
\startdata 
24 & 2\tablenotemark{*} & 0.6 & $<$0.08  & $<$10.9      & 0.8 & $<$4.60 \\
34 & 4\tablenotemark{*} & 3.0 & $<$0.2   & $<$27.2      & 0.6 & $<$5.21 \\
34 & 0                  & 1.8 & 0.2-1.0  & 27.2-135.9   & 0.4 & 5.21-6.26 \\
44 & 0                  & 1.6 & 1.0-8.0  & 135.9-1087.0 & 0.5 & 6.26-7.62 \\
\enddata
\tablenotetext{*}{Only one binary confirmed spectroscopically (PPl~15).    
CFHT-Pl-12, CFHT-Pl-16, and HHJ~6  are  
inferred to be binaries with nearly identical components because of their position in Figure~2, but they are not resolved with HST or AO images. 
q$_{\rm min}$ is the average minimum mass ratio (M$_2$/M$_1$) for a 
binary to be detected with our data.}
\end{deluxetable}

\clearpage

\centerline{\bf Figure Captions:}

\figcaption[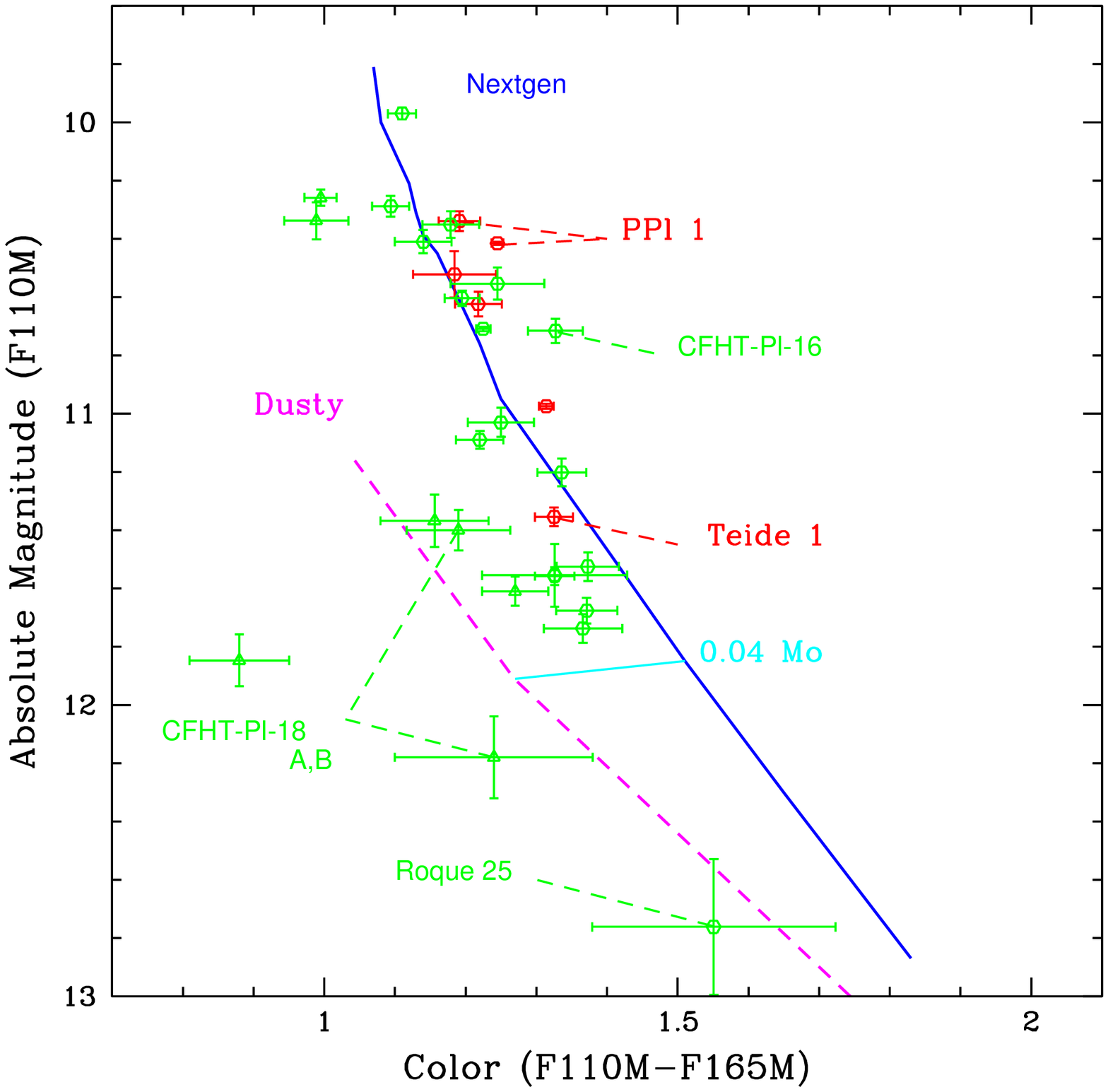]{\label{fig1} A NICMOS color-magnitude diagram. 
Lithium-confirmed Pleiades members are plotted with red color. The location of some 
important objects are indicated. PPl~1 has two 
positions in the plot because it is variable. 
We have used a Pleiades distance of 125 pc for calculating absolute magnitudes.
 The solid line (blue) is the 120 Myr isochrone for dust-free Nextgen model atmospheres. 
The dashed line (magenta) is an isochrone for the same age but Dusty model atmosphere. 
The cyan horizontal line joins the points corresponding to a mass of 0.04~M$_\odot$ for 
Nextgen and Dusty models.}

\figcaption[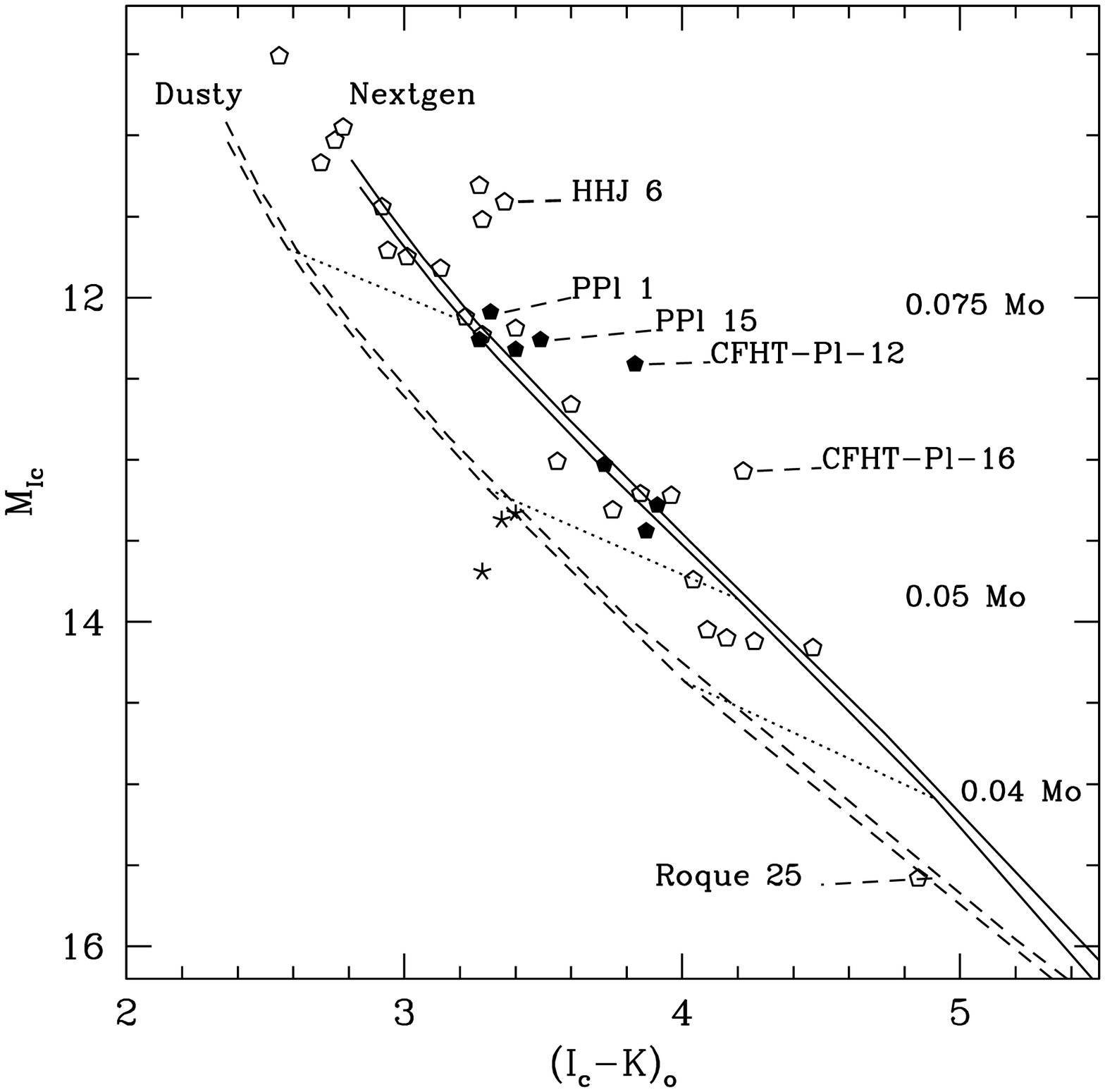]{\label{fig2} A color-magnitude diagram with the 
ground-based near-IR photometry.  Lithium confirmed BDs are denoted with filled pentagons. 
Objects that follow the cluster sequence are represented with emply pentagons. 
Objects that lie well below the sequence (likely non-members) are plotted as five pointed stars. 
The solid lines are the 100~Myr and 120~Myr isochrones for NextGen model atmospheres.  
The dashed lines are Dusty isochrones for the same ages. 
Dotted lines join points of the same mass (labelled in solar units) in different isochrones.}

\figcaption[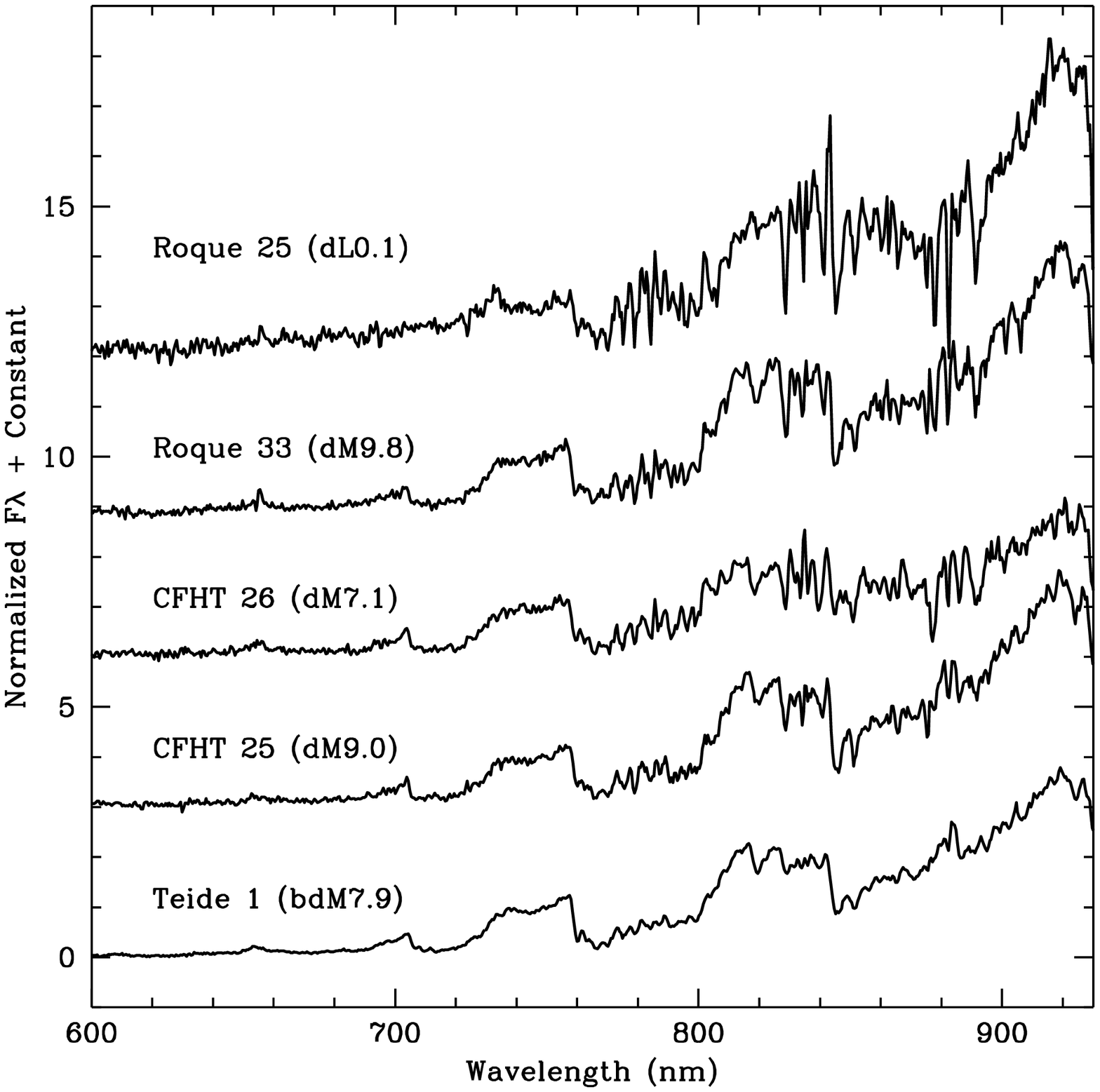]{\label{fig3} Keck LRIS low-resolution spectra 
of the coolest objects in our sample.}

\figcaption[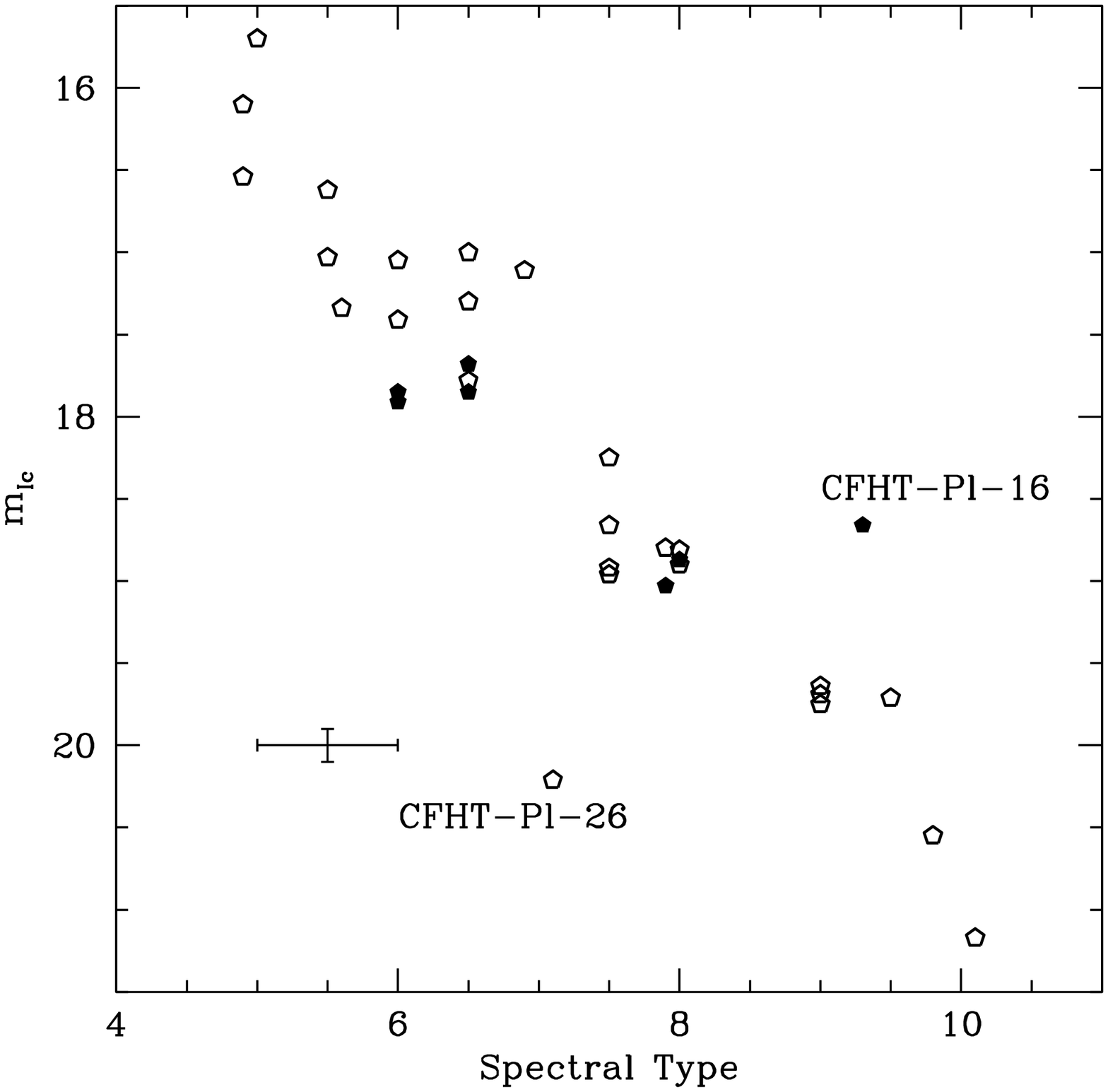]{\label{fig4} Spectral Types of Pleiades BD candidates versus 
observed I-band magnitudes. Filled symbols represent objects with lithium detections.}

\figcaption[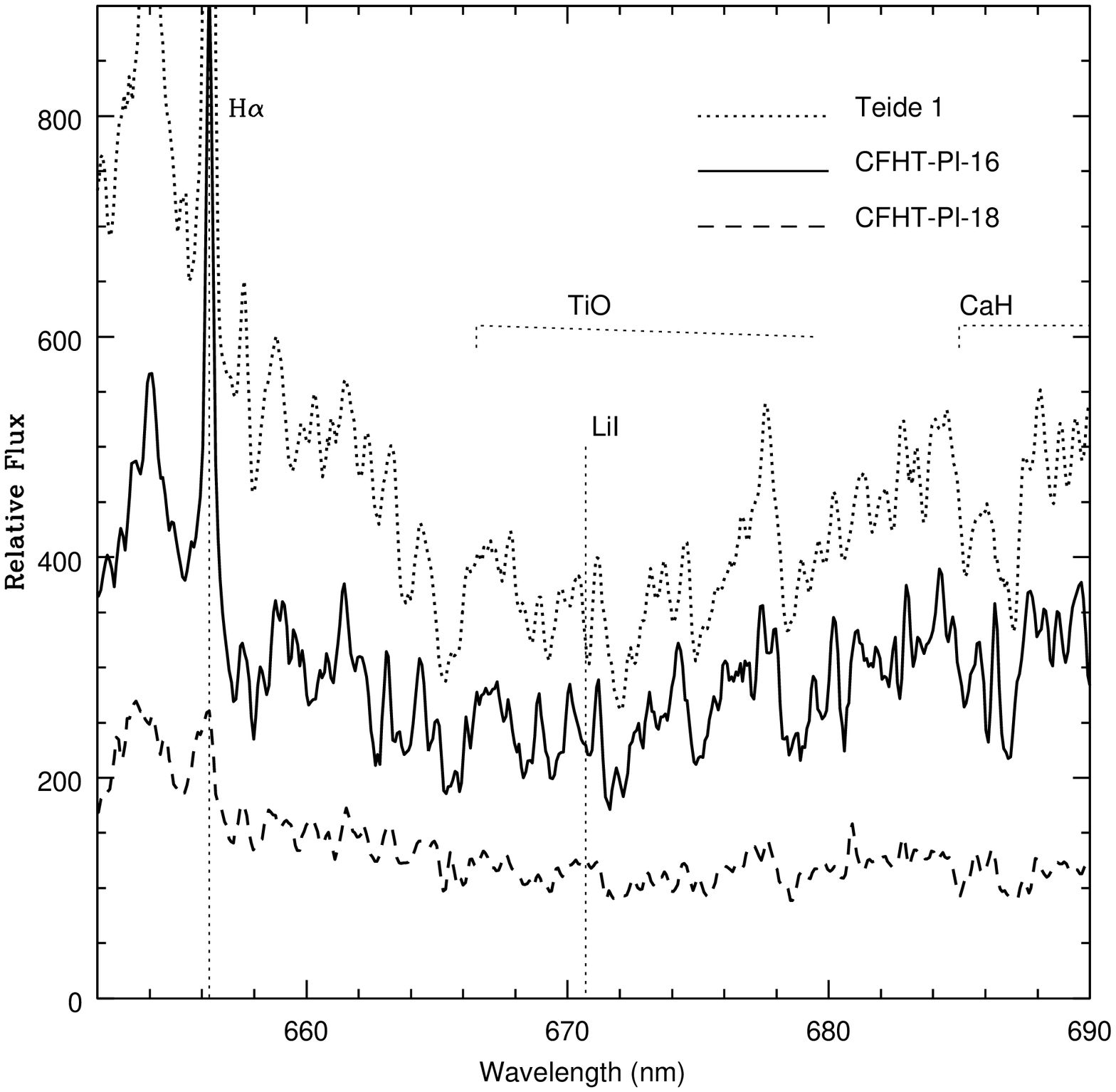]{\label{fig5} Keck LRIS mid-resolution spectra 
of CFHT-Pl-16 and 18. The spectrum of Teide~1 (Rebolo et al. 1996) is also shown 
for comparison. We have applied a boxcar smoothing of 3 pixels to all the spectra. 
H$_\alpha$ is seen in emission, and the LiI resonance line at 670.8~nm is detected 
in CFHT-Pl-16 but not in CFHT-Pl-18.}

\figcaption[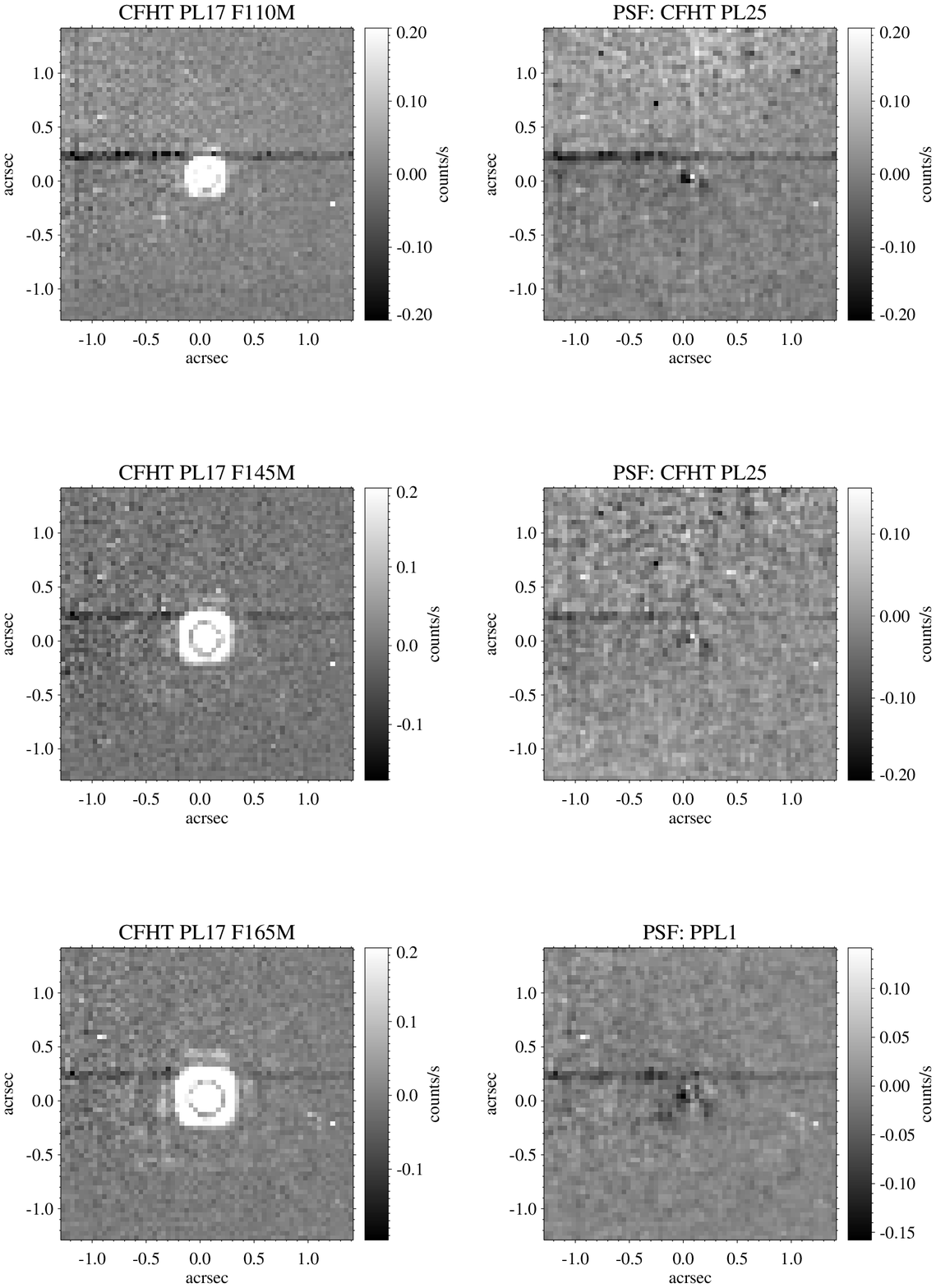]{\label{fig6} Reduced HST/NIC1 images of CFHT-Pl~17 
before and after PSF subtraction. The name of the object used for the subtraction is given on the right-hand side.}

\figcaption[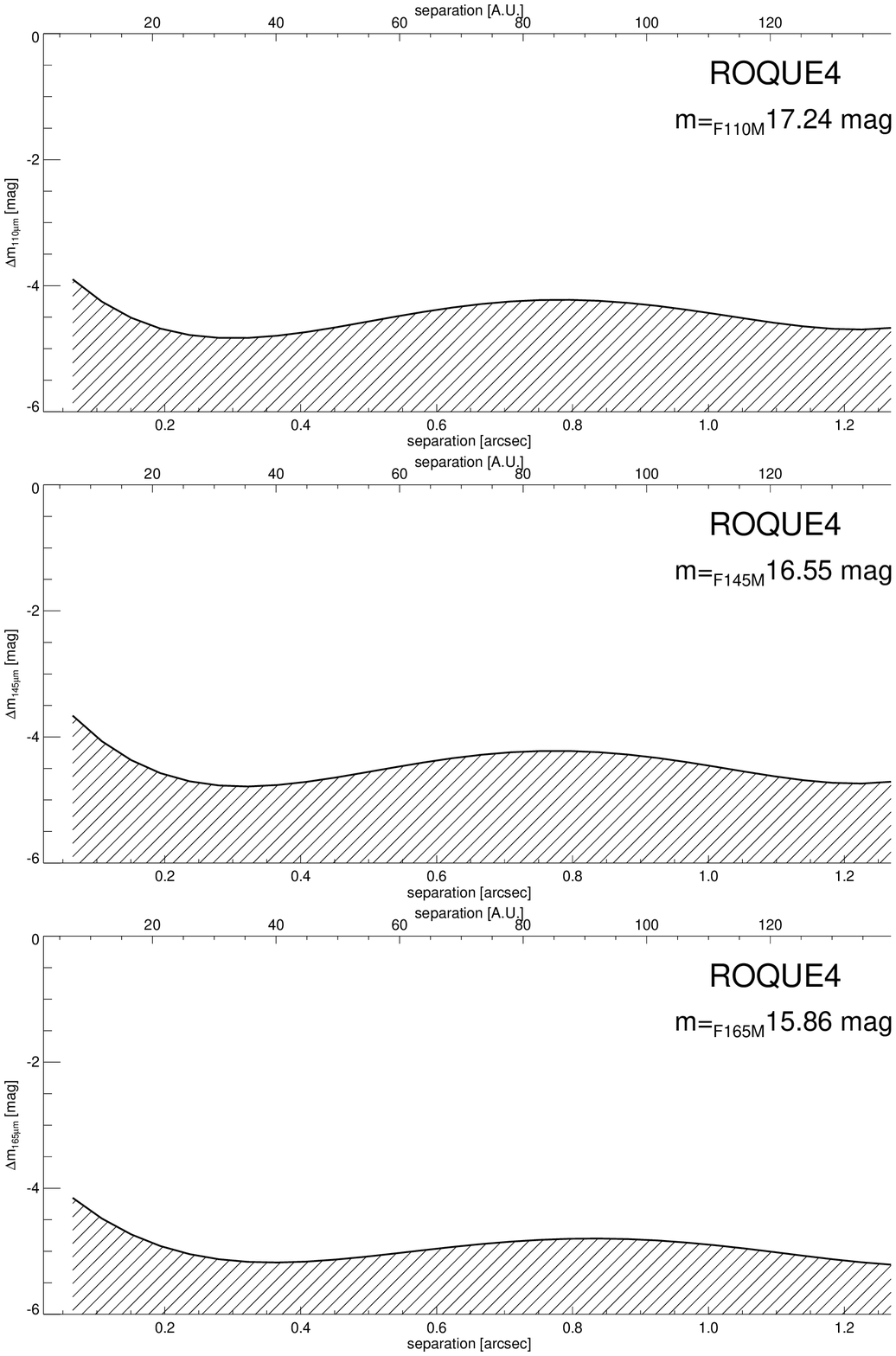]{\label{fig7} Examples of radially averaged 
sensitivity limits to the presence of companions after PSF subtraction.}

\figcaption[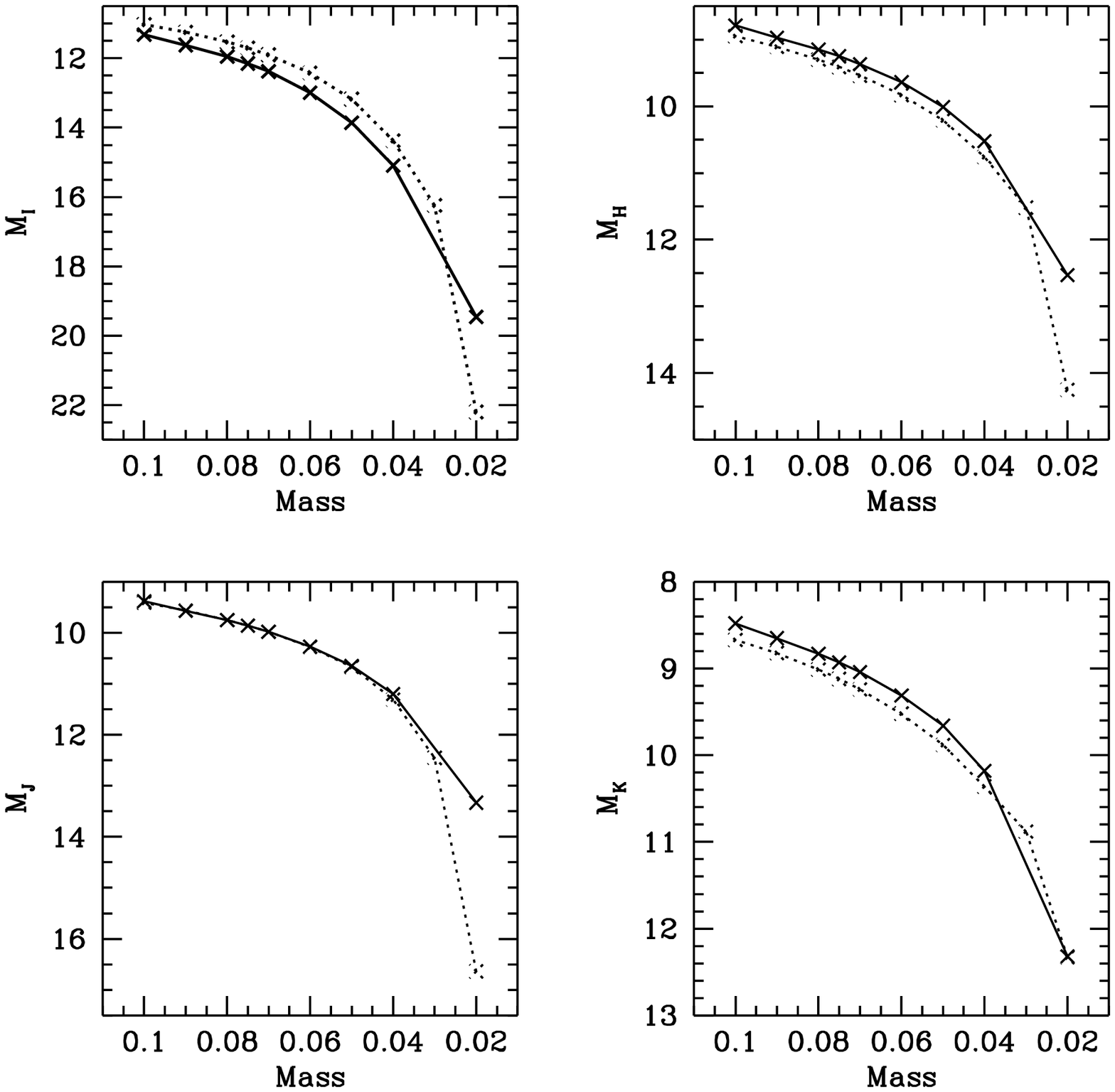]{\label{fig8} Mass-Luminosity relationships in different NIR filters 
given by Nextgen (solid line) and Dusty (dotted line) models for an age of 120~Myr.}

\clearpage

\plotone{emartin.fig1.ps}

\plotone{emartin.fig2.ps}

\plotone{emartin.fig3.ps}

\plotone{emartin.fig4.ps}

\plotone{emartin.fig5.ps}

\plotone{emartin.fig6.ps}

\plotone{emartin.fig7.ps}

\plotone{emartin.fig8.ps}

\end{document}